# Origin of ion bombardment induced Tb oxidation in Tb/Co multilayers

D. Kiphart*[a], M. Krupiński[b], M. Mitura-Nowak[b], P.P. Michałowski[c], M. Kowacz[a], M. Schmidt[a], F. Stobiecki[a], G.D. Chaves-O'Flynn[a], P. Kuświk*[a]

[a]*Institute of Molecular Physics, Polish Academy of Sciences, ul. M. Smoluchowskiego 17, 60-179 Poznań, Poland*

[b]*Institute of Nuclear Physics Polish Academy of Sciences, ul. Radzikowskiego 152, 31-342 Kraków, Poland*

[c]*Łukasiewicz Research Network—Institute of Microelectronics and Photonics, al. Lotników 32/46, 02-668 Warszawa, Poland*

*Corresponding authors: dkiphart@ifmpan.poznan.pl and kuswik@ifmpan.poznan.pl

**Abstract**

Ion bombardment is currently an active area of research for patterning rare earth/transition metal ferrimagnetic thin films because the magnetic properties are extremely sensitive to changes in the constituent sublattices. It has previously been shown that ion bombardment can be used to deliberately reduce the contribution of the rare earth sublattice in rare earth/transition metal ferrimagnets by selective oxidation. However, the exact mechanism by which oxidation occurs remains an outstanding question. We show that the defects introduced by ion bombardment of Tb/Co multilayers using different ion species with projected range (i.e., 10 keV $He^+$, 15 keV $O^+$, and 30 keV $Ga^+$) create easy diffusion paths for oxygen to penetrate the system. The choice of ion species and fluence enables the effective composition of the films to be tailored by reducing the amount of magnetically-active Tb.

**Keywords:** ferrimagnetic films, ion bombardment, oxidation, spin reorientation transition, compensation composition, thin films, multilayers

## 1. Introduction

Ion bombardment (IB) has been an effective tool for modifying ferromagnetic thin films and multilayers for many years because it allows for flexible adjustment of magnetic properties [1-16]. Due to the attractiveness of ferrimagnetic materials for spintronic applications [17], IB has also recently been an active area of research for patterning ferrimagnetic alloy thin films or multilayers [18-21]. In ferrimagnetic layered systems, the presence of two magnetic sublattices provides an opportunity to precisely modify the magnetic properties over a wide range. Both for ferromagnetic and ferrimagnetic systems, an appropriate choice of ion, energy ($E_{ion}$), and dose ($D_{ion}$), makes it possible to induce structural changes and enable control of magnetic properties e.g., magnetic anisotropy [4-6, 22], coercivity [7], and magnetic domination in the ferrimagnetic system [18-20]. This type of patterning offers high resolution and reproducibility at different length scales, including the nanoscale [8, 23]. This makes ion bombardment ideal for creating magnetic patterns for various applications, from data storage and computing to sensing and energy technologies. The ability to engineer and control magnetic patterns opens exciting opportunities for the development of modern technologies based on magnetic materials.

Recently, in the field of IB modification of magnetic layered systems, most of the scientific reports focused on the role of intermixing at the interface [9, 24], which leads to a reduction of the surface contribution to the effective anisotropy and therefore weakens perpendicular magnetic anisotropy (PMA) [10, 25]. On the other hand, due to intermixing at the interface, the magnetization can be reduced and, as a consequence, the effective anisotropy supporting PMA increases [11]. It was also demonstrated that intermixing is responsible for the presence of the chemical phases or strains that support PMA [12, 26]. It is also known that IB causes a reduction of orange peel coupling by roughness modification at the interface [14] and for very thin spacer supports ferromagnetic coupling due to pinhole formations [15, 16]. All of these interpretations are mainly based on the ballistic mixing process; however, it has recently been shown that IB with noble gases (e.g., $Ar^+$) can influence oxidation of metal thin-film systems [27, 28]. It was demonstrated that the oxygen chemisorption process depends on the specific metal, the energy and dose of bombarded ions, as well as oxygen exposure [28]. An increase in diffusivity has been proposed as a probable mechanism for oxidation enhancement after IB [27]. In the RE-TM (RE = rare earth, TM = transition metal) ferrimagnetic system, IB can also enhance oxidation. The added oxygen in this system selectively deactivates the magnetic moment of the RE sublattice because RE metals have a much greater susceptibility to oxidation than TM [29-31]. Consequently, the magnetic properties are modified, in particular magnetic anisotropy, coercivity, and the RE concentration corresponding to magnetic compensation of the RE and TM sublattices [18-20, 32]. Thus, IB can be performed locally for magnetic patterning of RE-TM alloy films or for RE/TM multilayers [18, 20, 21].

Here we focus on ion bombardment of Tb/Co multilayers (MLs) using different types of ions ($Ga^+$, $O^+$, $He^+$) in the energy range of 10-30 keV. We show that IB creates defects in the form of easy diffusion paths for oxygen penetration based on results from magneto-optic magnetometry, secondary ion mass spectrometry (SIMS), and electrical measurements and supported by Monte Carlo simulations. These results are well correlated with the number of generated vacancies that promote diffusion of oxygen to the ferrimagnetic system, which appears after irradiation.

## 2. Experiment

The samples were deposited at room temperature (RT) on naturally oxidized Si(100) substrates via magnetron sputtering. Complete details about the samples and the fabrication method can be found in the references [18, 33]. They consisted of Tb/Co multilayers with composition Ti-4 nm/Au-30 nm/(Tb-$t_{Tb}$/Co-0.66 nm)$_6$ /Au-5 nm, where $t_{Tb}$ was either a wedge from 0-2 nm or a uniform thickness, as described in each of the following sections and shown schematically in Figure 1.

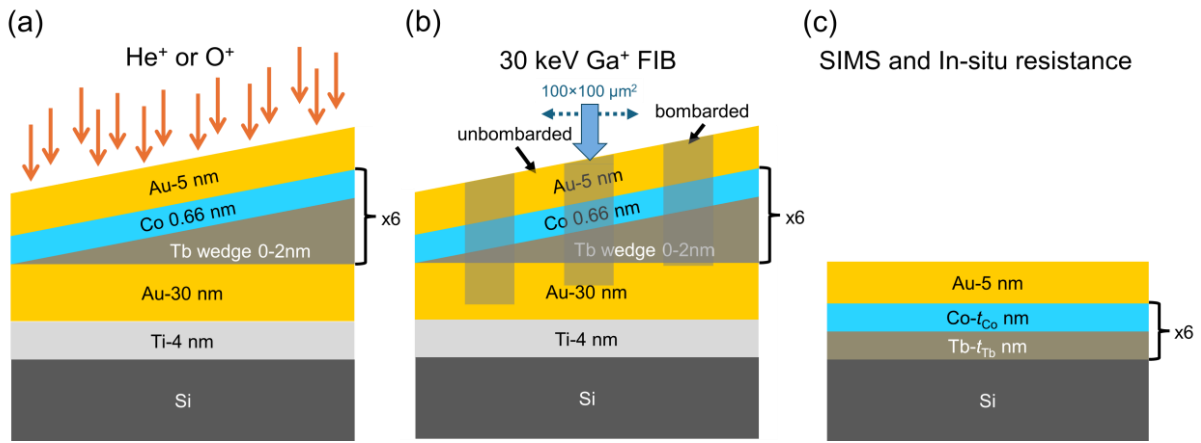

Figure 1. Schematic diagrams of the wedge samples used for $He^+$ and $O^+$ IB (a) and for $Ga^+$ FIB (b) and uniform samples for SIMS and in-situ resistance measurements (c). The thicknesses of the Co and Tb sublayers ($t_{Co}$ and $t_{Tb}$ respectively) are given in the text. The wedge samples for $He^+$ and $O^+$ IB were bombarded with stripes of uniform dose along the direction of the wedge. The wedge sample for $Ga^+$ FIB was bombarded in 100×100 µm² square regions with uniform dose along the direction of the wedge. Each square region was spaced 500 µm center-to-center, giving 33 squares per dose.

The ion bombardment using $He^+$ and $O^+$ was performed using an Eaton NV-3204 implantation system with an ion energy of 10 keV for $He^+$ and 15 keV for $O^+$. The vacuum pressure in the irradiation chamber was $7\times10^{-5}$ mbar for He$^+$ and $8.4\times10^{-5}$ mbar for O$^+$. Wedge samples ($0 \leq t_{Tb} \leq 2$ nm) were chosen to study the impact of ion bombardment versus composition. The films were bombarded with stripes of uniform dose along the Tb wedge by using a mask. The width of each stripe was approximately 2 mm wide and the stripes were separated by 2 mm for unmodified reference areas. The samples were deposited on $15\times19$ mm$^2$ substrates and therefore three doses could fit on a single sample. One sample was bombarded with He$^+$ ions ($D_{He}$ = 1, 2, 3 $\times10^{15}$ He$^+$/cm$^2$) and two samples were bombarded with O$^+$ ions ($D_O$ = 0.1, 0.3, 0.7, 1, 3, 7 $\times10^{15}$ O$^+$/cm$^2$).

The magnetic properties of the ML were measured at RT using polar magneto-optical Kerr effect (P-MOKE) using a 655 nm wavelength laser with a spot size of approximately 0.2 mm. The changes in magnetic properties as a function $t_{Tb}$ and $D_{ion}$ were investigated by moving the sample relative to the stationary laser beam. The magnetic properties of the samples were measured in the as-deposited state and after ion bombardment in both the modified regions and unmodified reference areas.

For experiments with 30 keV Ga$^+$, we used a focused ion beam (FIB) (FEI Helios Nanolab 660) workstation. The IB was performed along the wedge sample ($0 \leq t_{Tb} \leq 2$ nm) on a series of 33 squares with size $100\times100$ µm$^2$ patterned with a uniform ion dose. The distance between centers of the squares was 500 µm. We repeated this procedure with the range $D_{Ga}$ = 1-15$\times10^{13}$ Ga$^+$/cm$^2$ to investigate the influence of the different doses on the magnetic properties of Tb/Co MLs. In this case, the magnetic properties were measured using a P-MOKE microscope (evico magnetics GmbH).

The impact of Ga$^+$ FIB on the structural morphology of the ML system was investigated by obtaining composition depth profiles of the sample in the as-deposited and bombarded states by secondary ion mass spectrometry using a CAMECA SC Ultra instrument [18]. This sample had a composition of (Tb-0.8 nm/Co-0.8 nm)$_6$/Au-5 nm (Fig. 1c) and was patterned using Ga$^+$ FIB with a uniform fluence of $8\times10^{15}$ Ga$^+$/cm$^2$ and different energies $E_{Ga}$ = 8, 16, 30 keV. The thickness of the Co sublayer was slightly increased to provide better resolution in the SIMS measurements and improve the distinction between the different layers and as well as the distribution of oxygen after ion bombardment.

In order to gain insight into when the oxidation process occurs, the resistance of a sample with composition (Tb-0.82 nm/Co-0.66 nm)$_6$/Au-5 nm [Fig.1(c)], was measured before, during, and after IB with 10 keV He$^+$. The electrical contacts were connected to the sample using silver paste and were covered by tape to protect them from interaction with ions during IB. Each data point was calculated from two measurements with opposite direction of the current flow to reduce measurement noise. The sample was inserted into the implanter, where both the beam and the sample were stationary. The pressure during irradiation was about $5\times10^{-5}$ mbar. The sample remained in the chamber after bombardment in a vacuum of $5\times10^{-6}$ mbar for approximately 22 hours. Afterwards, the chamber was vented and the resistance continued to be measured for the next 3 days under ambient pressure.

Monte Carlo simulations were performed using the TRIDYN code [34] to obtain a quantitative estimate of the microstructural damage caused by IB (e.g., the distribution of introduced vacancies, interlayer mixing, and ion implantation). The simulations were performed for a system with an initial composition of Ti-4 nm/Au-30 nm/(Tb-1 nm/Co-1 nm)$_6$/Au-5 nm and bombarded with 10 keV He$^+$, 15 keV O$^+$, and 30 keV Ga$^+$ with doses ranging from $D_{ion}$ = 0.02$\times10^{15}$ to $D_{ion}$ = 1$\times10^{15}$ ions/cm$^2$.

## 3. Results

To determine the influence of IB on the magnetic properties of Tb/Co MLs, we have chosen helium, oxygen, and gallium. For He$^+$ and O$^+$ we performed IB on Tb/Co multilayers (MLs) with constant $D_{ion}$ along the gradient of $t_{Tb}$ for varying $D_{ion}$. In the experiments with Ga$^+$, we performed bombardment using FIB. In this technique, the bombarded surface area is severely constrained, so in this case we uniformly bombarded areas of $100\times100$ µm$^2$ with $1\times10^{13} \leq D_{Ga} \leq 15\times10^{13}$ Ga$^+$/cm$^2$ at different thicknesses of Tb in the (Co/Tb-wedge) sample. This allowed us to determine thicknesses corresponding to compensation of the Co and Tb sublattices ($t_{Tb}^{comp}$) and spin reorientation transition ($t_{Tb}^{SRT}$), (from in-plane to out-of-plane anisotropy) based on magnetization reversal measurements along the wedge for all ions. From collected hysteresis loops, we determined $H_C$ and $\varphi_R/\varphi_S$ ($\varphi_R$ and $\varphi_S$ are magneto-optical Kerr signal at remanence and saturation, respectively) as a function of $t_{Tb}$ (Fig. 2). For all ions, the $t_{Tb}^{comp}$ [Fig. 2(a,b,c)] and the $t_{Tb}^{SRT}$ [Fig. 2(d,e,f)] are shifted to higher $t_{Tb}$ with increasing $D_{ion}$, which shows the same trend as was found previously [18, 19]. Note that the shift for Ga$^+$ appears in a dose range which is two orders of magnitude smaller than that of He$^+$; and one order of magnitude smaller than that of O$^+$. This means that fewer Ga$^+$ ions are necessary to achieve the same shift as for O$^+$ or He$^+$. However, for all ions, we found a linear dependence between $D_{ion}$ and the shift of $t_{Tb}^{comp}$ [$\Delta t_{Tb}^{comp} = t_{Tb}^{comp}(D_{ion}) - t_{Tb}^{comp}(D_{ion}=0)$] as well as the shift of $t_{Tb}^{SRT}$ [$\Delta t_{Tb}^{SRT} = t_{Tb}^{SRT}(D_{ion}) - t_{Tb}^{SRT}(D_{ion}=0)$] (Fig. 3).

Taking into account that PMA in Tb/Co MLs has two main sources: interfacial [35] and single ion anisotropy [36], one can expect stronger modification of $t_{Tb}^{SRT}$ than $t_{Tb}^{comp}$ due to the reduced contribution from Tb to single ion anisotropy (Tb deactivation by oxygen) as well as mixing at the interfaces. This is not the case for O$^+$ and He$^+$, as weaker changes were found for $t_{Tb}^{SRT}$ than $t_{Tb}^{comp}$. To analyze these results, we determined the slopes ($S$) of the linear fits for $\Delta t_{Tb}^{comp}(D_{ion})$ and $\Delta t_{Tb}^{SRT}(D_{ion})$ dependencies ($S^{SRT}$ and $S^{comp}$, respectively). Taking the ratio $R=(S^{SRT}/S^{comp})$ for each ion species, we found that $R$ is almost identical for He$^+$ ($R = 0.64 \pm 0.01$) and O$^+$ ($R = 0.67 \pm 0.02$). This means that the impact of IB is less pronounced for $t_{Tb}^{SRT}$ than for $t_{Tb}^{comp}$, since the $t_{Tb}^{SRT}$ is always much smaller than $t_{Tb}^{comp}$. This suggests that the amount of available Tb for oxidation is crucial for the change in magnetic properties and that this change is related to the reduction of the amount of magnetically active Tb in the system. Therefore, $\Delta t_{Tb}^{comp}$ and $\Delta t_{Tb}^{SRT}$ can be interpreted as an effective oxidation thickness of the Tb sublayers. As for Ga$^+$, we found that the $t_{Tb}^{SRT}$ and $t_{Tb}^{comp}$ are changing at almost the same rate, which gives $R=1.01 \pm 0.02$. In the case for Ga$^+$, we cannot exclude that the shift of $t_{Tb}^{SRT}$ is also related to the modification of the interface, which causes a significant reduction of the surface contribution to the effective anisotropy which supports PMA. This statement is supported by the stronger ballistic mixing effect found on the concentration profiles of Tb, Co and Au after Ga$^+$ bombardment than for He$^+$ or O$^+$ for the same $D_{ion}$ (see Supplementary Fig. S2). Therefore, an effective oxidation thickness of the Tb sublayers for Ga$^+$ can be determined only from $\Delta t_{Tb}^{comp}$.

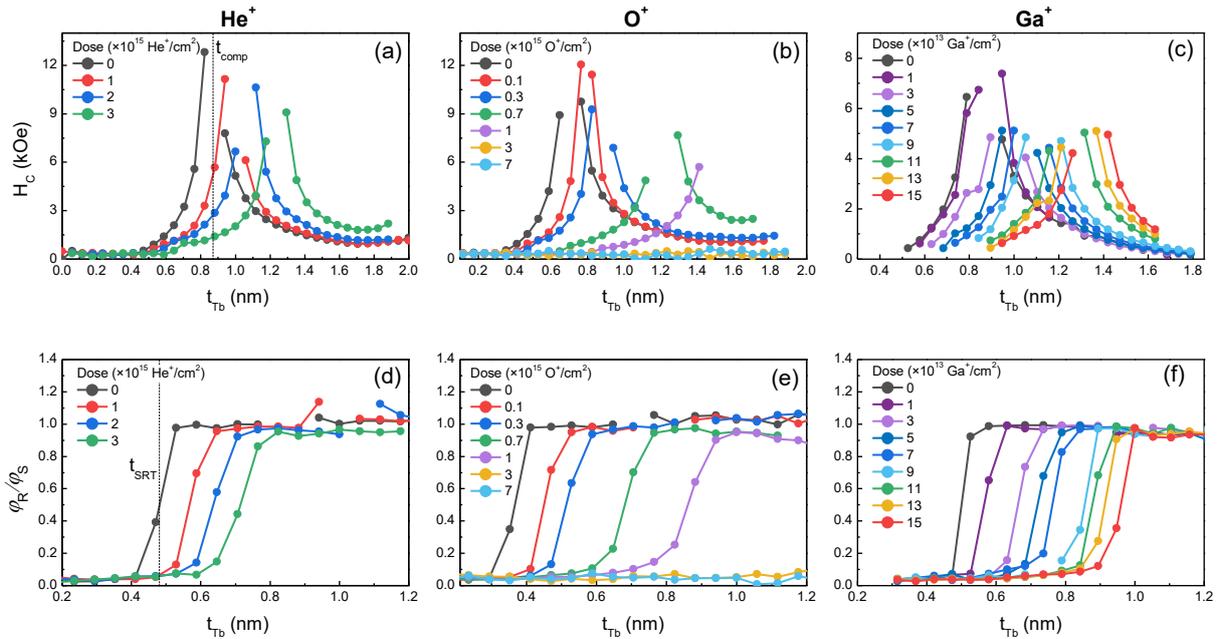

Figure 2. Coercivity ($H_C$) (a, b, c), ratio of Kerr signal at remanence ($\varphi_R$) and at saturation ($\varphi_S$) (d, e, f) versus thickness of Tb sublayers ($t_{Tb}$) for different doses of ion bombardment with 10 keV He$^+$ (a, d),

15 keV O$^+$ (b, e), and 30 keV Ga$^+$ (c, f). The dashed lines in (a) and (b) indicate the definition of $t_{Tb}^{comp}$ and $t_{Tb}^{SRT}$, respectively, for $D_{ion}$=0. The samples were MLs with composition: Ti-4 nm/Au-30 nm/(Tb-wedge 0-2 nm/Co-0.66 nm)/Au-5. Details of the IB procedures can be found in the Experiment section.

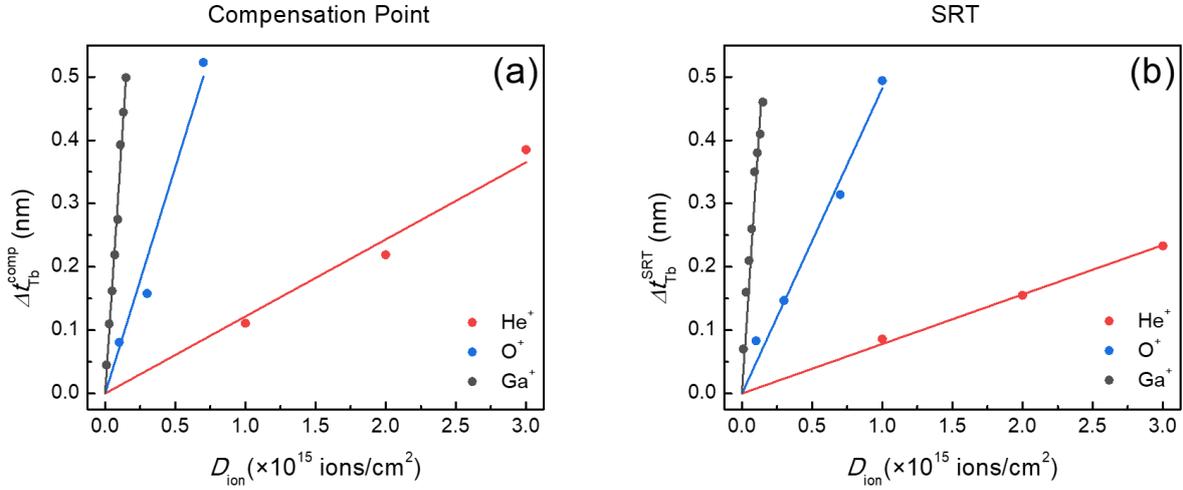

*Figure 3. The change in the thickness of the Tb sublayer for which the ML system is magnetically compensated ($\Delta t_{Tb}^{comp}$) (a) and spin reorientation transition ($\Delta t_{Tb}^{SRT}$) (b) as a function $D_{ion}$ for 10 keV He$^+$, and 15 keV O$^+$ and 30 keV Ga$^+$ IB. The lines denote the linear fit of the data points with the intercept fixed at the origin, as by definition the effective Tb thickness should be unchanged for $D_{ion} = 0$.*

Recently, it was demonstrated that oxygen can be found inside Tb/Co ML systems bombarded by He$^+$ [18, 20, 32]. To verify if the same effect appears for Ga$^+$, we measured the depth profile in the as-deposited state and after bombardment with different Ga$^+$ energy ($E_{ion}$ = 8, 16, and 30 keV) (Fig. 4). The SIMS data reveal that oxygen is present and that the maximum depth depends strongly on ion energy. For 8, 16, and 30 keV the oxygen was detected up to the ~15, 18, and 23 nm, respectively. This result can be understood by taking into consideration that the ion range strongly depends on the ion energy, therefore sample damage can be expected only where ions can reach. This means that oxygen can be found deeper for higher ion energy. It is also important that for all energies the position of the oxygen peaks correlates well with the peak position for Tb. This means that oxygen is preferentially located in the Tb sublayer and, as a result, the contribution to the ferrimagnetic properties of Tb decreases similarly to that found for He$^+$ [18]. It should be noted that selective oxidation of metallic multilayers was also observed for other systems for which this effect was attributed to different oxidation potentials of the metals forming the multilayer system [37].

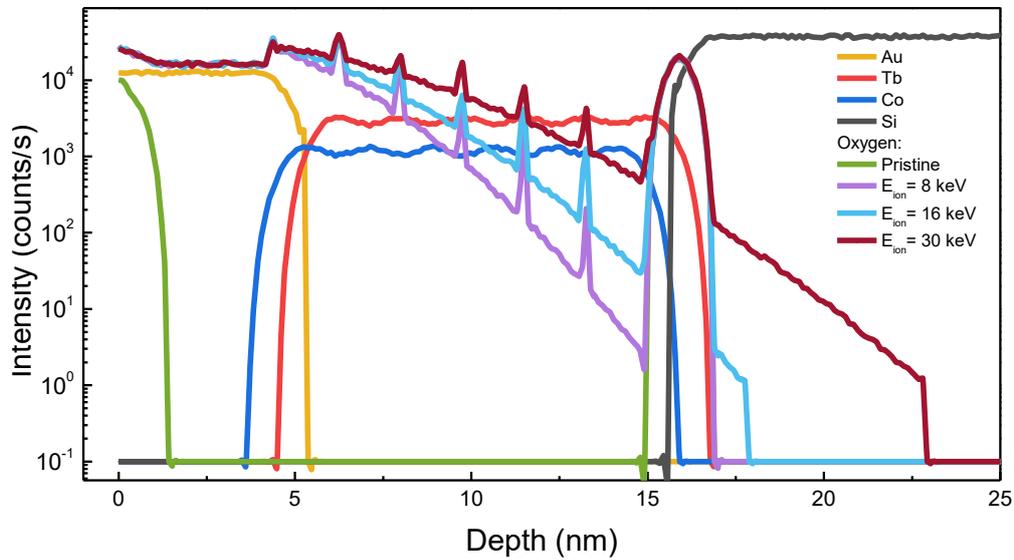

*Figure 4. SIMS depth profile for (Tb-0.8 nm/Co-0.8 nm)$_6$/Au-5 nm in the pristine state and after Ga$^+$ FIB using different $E_{ion}$ with constant $D_{Ga}=8\times10^{13}$ Ga$^+$/cm$^2$.*

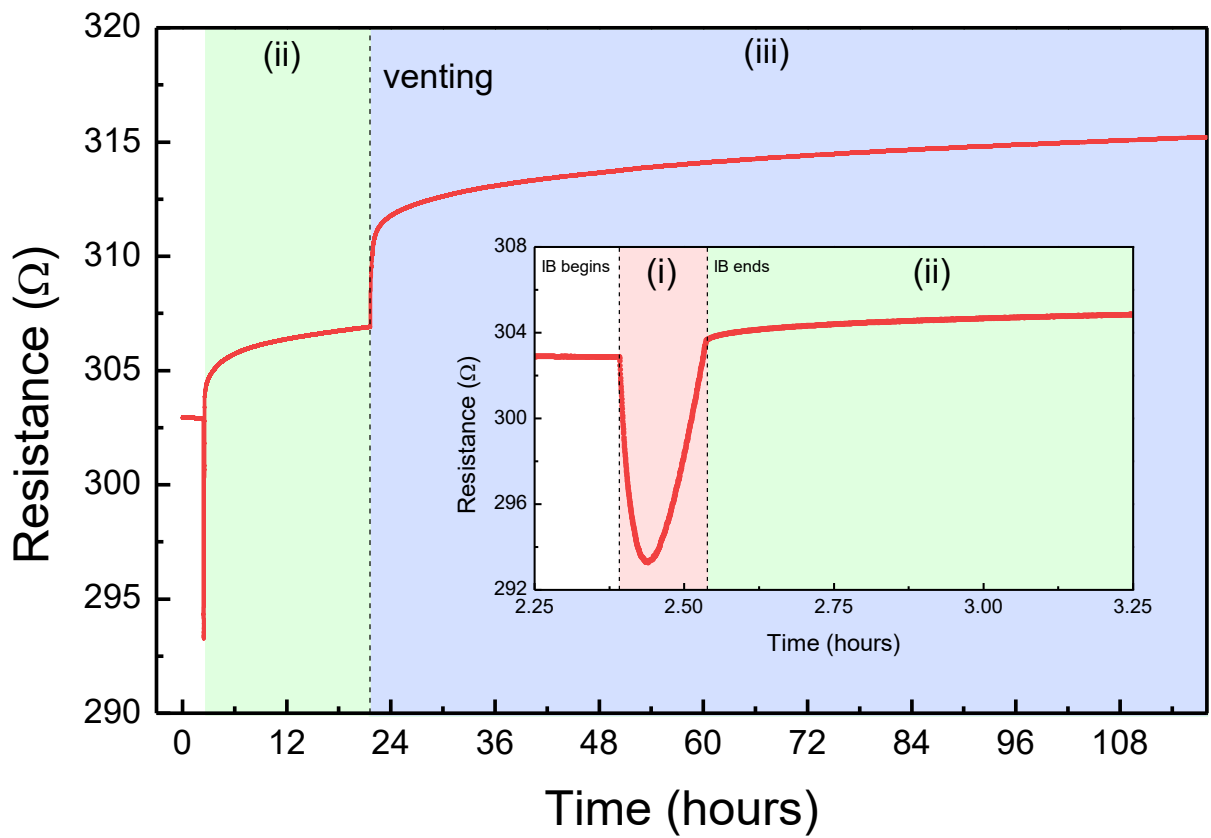

*Figure 5. Resistance versus time measured in-situ before, during and after 10 keV He$^+$ IB with $D_{He}=10^{15}$ He$^+$/cm$^2$. The inset shows the time interval during IB. The dashed lines indicate the start and end of IB and the start of venting and the shaded regions denote the three time intervals referred to in the text.*

To identify the origin of the oxygen present in the samples we decided to measure the resistivity of the Tb/Co ML before, during, and after the IB procedure while keeping the sample under vacuum ($p_0 = 5\times10^{-5}$ mbar). We also measured the resistivity after venting the chamber. The results are presented in Fig. 5 where we highlight three time intervals: (i) during IB, (ii) after IB but in vacuum, and (iii) after venting. During that experiment, we also measured the sample temperature, which was around 23 °C for the whole experiment and did not change by more than 2 °C. Changes in resistance during the bombardment process [interval (i), shown in Fig. 5 inset] show a drop and recovery of the resistance to a value barely higher than that at the start of the process. Such non-monotonic changes were previously observed in thin films during ion bombardment [38, 39] and it can be caused by various factors. For example, the decrease of the resistance might be due to occluded gas release by ions [39]. Additionally, ion-irradiation induces stress relaxation in the metallic multilayer [40, 41] and in amorphous materials [42], which should result in a decrease in the resistance. After the initial fall, the increase of resistance might be due to defect formation as well as oxidation [43 and references therein]. The detailed mechanisms of resistance changes during IB [interval (i)] are complex, which were described in detail in refs. [44, 43 and references contained therein].

Immediately after the end of the IB process [interval (ii)], the resistance of the sample began to increase, which can be interpreted as oxidation of the ML system [43 and references therein]. This process started fast and slowed down after a few hours; nevertheless, the resistance continued to slowly increase for next several hours. This can be interpreted as a continuation of the oxidation process due to residuals in the chamber that contain oxygen (e.g., $H_2O$, $O_2$). At the beginning of the oxidation process, vacancies and other ion-generated defects (e.g., line defects) seem to play a key role, which is visible by the rapid increase in sample resistance several seconds after the end of the irradiation process.

After approximately 22 hours, the chamber was vented with air [interval (iii)], which caused another significant increase in resistance, which indicates that the oxidation process accelerated again and slowly saturated over time.

We also performed a similar IB procedure with much higher pressure in the chamber ($2\times10^{-4}$ mbar). The greater amount of residual gas and oxygen resulted in a larger increase in resistance after the first irradiation cycle. All these results show, together with SIMS data, that the oxygen strongly penetrates Tb/Co MLs after irradiation [intervals (ii) and (iii)]. However, we cannot exclude that oxygen diffusion may begin during IB.

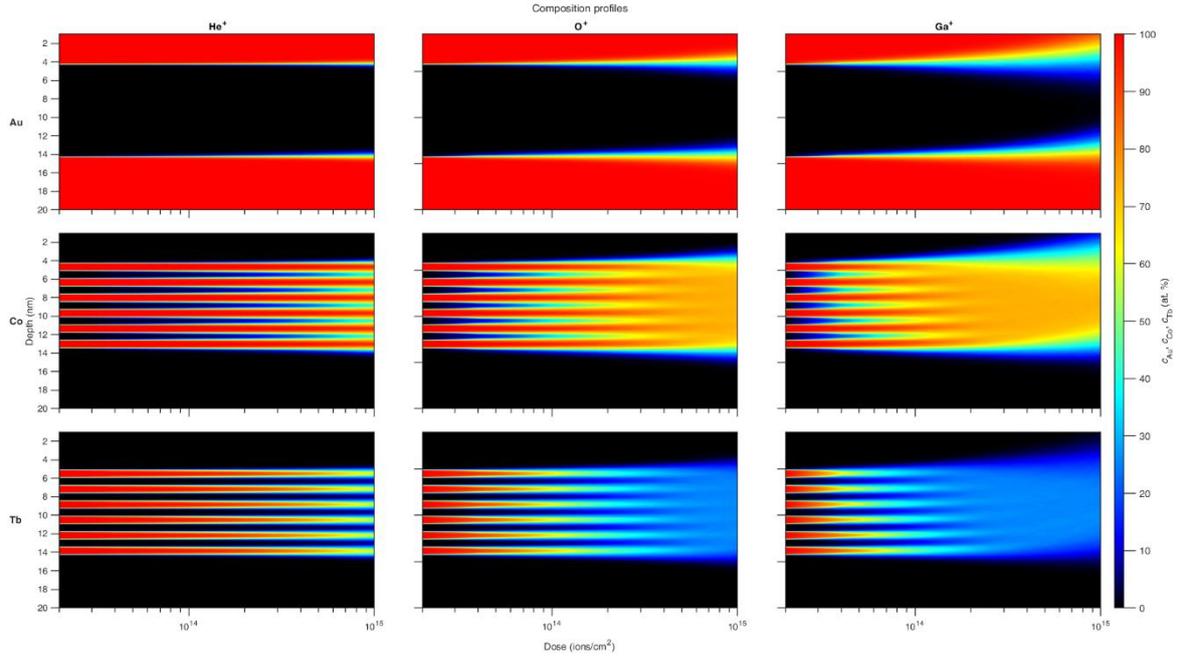

*Figure 6. Composition depth profiles obtained from TRIDYN simulations for the top 20 nm of a Ti-4 nm/Au-30 nm/(Tb-1 nm/Co-1 nm)/Au-5 nm ML. Each column shows the change in composition versus $D_{ion}$ for 10 keV $He^+$, 15 keV $O^+$ and 30 keV $Ga^+$. Each row shows the concentration of each element separately in atomic percent ($c_{Au}$, $c_{Co}$, and $c_{Tb}$, respectively).*

To understand why the oxygen can penetrate MLs after bombardment we used TRIDYN to simulate the material distribution for $He^+$, $O^+$, and $Ga^+$ ions (Fig. 6) and to calculate vacancy density (Fig. 7). From the material distributions simulated, we see that for a given dose $Ga^+$ exhibits the strongest intermixing. It should be emphasized that for the highest $D_{Ga}$ the mixing is so strong that the MLs become alloys. The IB causes significant displacements of all elements (Au, Co, Tb); however, the level of intermixing between each sublayer strongly depends on the ion mass. Considering that we use all ions in a similar ion energy range (tens of keV), these results can be connected mainly with the ion mass. The high kinetic energy is transferred to the material, which enables the relocation of the atoms even several lattice constants away from the initial position. This causes strong mixing between layers, which is a well-known effect responsible for reducing the interface contribution to the PMA in ferromagnetic layered systems [25]. On the other hand, IB creates a significant number of vacancies [45]. Based on the TRIDYN data we calculated the vacancy density up to 17 nm, which is the maximum depth where the magnetic material was found after ion induced mixing. It is clear from this data (Fig. 7), that the vacancy density is the lowest for $He^+$ and highest for $Ga^+$, and that the vacancy density depends linearly on the dose for all ions. Note that the same behavior was found for $\Delta t_{Tb}^{SRT}$ and $\Delta t_{Tb}^{comp}$ as can be seen in Fig. 3.

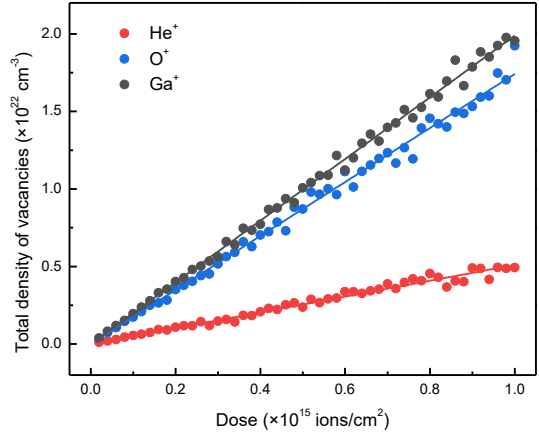

*Figure 7. Calculated density of vacancies introduced during IB as a function of $D_{ion}$ for 10 keV He$^+$, 15 keV O$^+$ and 30 keV Ga$^+$. The points are results from TRIDYN simulations and the lines are the results of linear fit of the simulated data with the intercept fixed at zero.*

## 4. Discussion

To start, we will discuss the origin of oxygen in MLs. The first direct observations of oxygen in Tb/Co MLs were performed for a bombarded sample using a home-built Penning ion source [46]. Even using ultra-pure He gas (5N) one can expect some residual oxygen, which can be ionized and implanted into the MLs. Since we found similar magnetic changes of the sample bombarded with 10 keV He$^+$ using an implantation system with mass separator (bending magnets), and in this case we can exclude the possibility of implantation with ions different to He$^+$, we conclude that this mechanism for oxidation is very unlikely. We also know, from SIMS measurements (Fig. 4), that for all Ga$^+$ energies oxygen is present in MLs and in the same way as He$^+$ causes the magnetic deactivation of Tb. In the FIB technique, ions are produced from a liquid metal ion source under a high vacuum ($p=3\times10^{-7}$ mbar) excluding the possibility of producing oxygen ions. These all indicate that the presence of oxygen in MLs is not related to implantation of O$^+$. In the case of IB with oxygen, we cannot exclude the effect of oxidation directly from the beam; however, since the TRIDYN simulations show that the amount of oxygen inside the layers coming from the beam does not exceed more than 0.6% for $D_O=10^{15}$ O$^+$/cm$^2$ (see SM Fig.S1) and the mean number of implanted oxygen atoms per ion is 0.219 (taking into account only top 17 nm of MLs), this small amount of implanted oxygen cannot explain why the shift of $t_{Tb}^{comp}$ and $t_{Tb}^{SRT}$ is larger for O$^+$ than for He$^+$. To show that we calculated number of Tb atoms which are oxidized per ion for each ion species based on the data from Fig. 3. On that figure, the slopes of the linear fits represent the approximate volume of Tb deactivated for every 10$^{15}$ ions. Taking this value and dividing by the atomic volume ($3.21\times10^{-23}$ cm$^3$/atom) and the dose factor 10$^{15}$, we estimated the number of Tb atoms deactivated per ion, as shown in Table 1. This data shows that nearly 2.45 Tb atoms are deactivated for each O$^+$ ion. This is ~10 times more than what is obtained from direct implantation. Therefore, another mechanism must be taken into consideration.

TABLE I. The calculated number of Tb atoms which are oxidized per ion for each ion species considered and Tb sublayer thicknesses close to the compensation point (CP) and SRT.

| Ion Species | 10 keV He$^+$ | 15 keV O$^+$ | 30 keV Ga$^+$ |
|---|---|---|---|
| Tb deactivated CP (atoms/ion) | 0.382 ± 0.02 | 2.23 ± 0.18 | 10.4 ± 0.19 |
| Tb deactivated SRT (atoms/ion) | 0.246 ± 0.01 | 1.51 ± 0.06 | -- |

A second possible mechanism, proposed earlier [18, 32], is the creation of easy diffusion paths, which enable the diffusion of oxygen deep into the layered system through them. This type of oxidation process is well known in the field of material science, since the creation of vacancies is triggered by the collision of ions with material atoms [45]. Our TRIDYN simulations show that IB with Ga$^+$, O$^+$, and He$^+$ generate 33, 29, and 9 vacancies/ion, respectively, and that the density of vacancies increases linearly with $D_{ion}$, which has the same trend as we found analyzing the $\Delta t_{Tb}^{SRT}(D_{ion})$ and $\Delta t_{Tb}^{comp}(D_{ion})$ (Fig. 3). These values show that the most vacancies are created for Ga$^+$, which explains why for this type of ion the dose necessary for the same modification of magnetic properties are the smallest (Table 1). On the other hand, He$^+$ generates much fewer vacancies, which means that IB has to be performed in a higher dose range to oxidize Tb/Co MLs (Table 1). It is worth noting that for very thin RE and TM sublayers (in the case of Tb-Co, where $t_{Co} = t_{Tb} \leq 1.5$ nm) in RE/TM multilayers, the magnetic properties are almost identical to those of RE-TM alloys [18 and references therein]. This similarity extends not only to the magnetic properties but also to the structural characteristics, as both systems exhibit an amorphous structure [35, 47]. Since almost all the data were analyzed for $t_{Co}$ and $t_{Tb}$ below 1.5 nm, we can assume that our system is also amorphous. Consequently, we cannot rule out additional effects that may promote oxygen diffusion in the multilayers following the IB process, as amorphous sputtered materials can contain grains and grain boundaries [48]. These effects may include, for example, the formation of anisotropic defects [49] and variations in diffusion coefficients between the grains and their boundaries.

We also found that $R$ for O$^+$ and He$^+$ bombardment has almost the same value. This shows that the Tb oxidation process is proportional to $t_{Tb}$. Assuming that the easy diffusion paths are oriented along the direction normal to the layer, the oxidation of Tb regions located near these paths will be more pronounced with increasing $t_{Tb}$, which strongly suggests that the vacancy creation, together with the preferential oxidation of Tb atoms, is the main process responsible for oxygen diffusion into the ML. This process occurs also for Ga$^+$ as we found a preferential distribution of oxygen in the Tb sublayers (Fig. 4). Therefore, the important question is where the oxygen comes from. Based on the resistance measurement, we found that the resistance increases after the bombardment which is interpreted as oxidation. This process is triggered by the residuals in the chamber, which contains oxygen. However, it cannot be ruled out that the oxygen diffusion process takes place from a thin layer adsorbed on the surface of the ML [50]. Importantly, removing the irradiated sample from the vacuum chamber into the atmosphere speeds up this process and ends it with stronger oxygen adsorption. This important observation confirms that the source of the oxygen inside the ML is not coming from the ion beam but it is mainly related to the diffusion of the oxygen after the bombardment process.

Oxygen diffusion through easy diffusion paths introduced by ion bombardment has not previously been considered in magnetic layered systems. In the case of ferrimagnetic layers containing rare earths, this process is strong as those elements are extremely easy to oxidize. Since ferromagnetic materials can also easily form oxides [51-53], the diffusion of oxygen into the ferromagnetic layers should also be considered when analyzing the influence of ion bombardment on magnetic properties in these systems.

## 5. Conclusions

In this study, we examine the effects of ion bombardment on Tb/Co multilayers using various ions ($Ga^+$, $O^+$, and $He^+$) within a low energy range (10-30 keV). The results of magneto-optic and SIMS measurements show that the change in magnetic properties is caused by the decrease in the amount of magnetically active Tb due to preferential oxidation. These results are well correlated with the results from Monte Carlo simulations, in which we found a linear dependence of the vacancies versus dose-similar to that found for the reduction in magnetically active Tb layer by oxidation. Most of the oxygen is introduced after venting the chamber, as shown by the electrical measurements. We have shown that ion bombardment induces defects that create easy diffusion paths for the oxygen penetration that takes place after irradiation.

**Acknowledgements**

This research was funded by National Science Centre Poland under OPUS funding Grant No. 2020/39/B/ST5/01915. For the purpose of Open Access, the author has applied a CC-BY public copyright licence to any Author Accepted Manuscript (AAM) version arising from this submission.

The data that support this study are available via the Zenodo repository, https://doi.org/10.5281/zenodo.14387846

**Author contributions**

**Daniel Kiphart:** Conceptualization, Methodology, Formal analysis, Investigation, Writing – review & editing. **Michał Krupiński:** Methodology, Investigation, Writing – review & editing. **Marzena Mitura-Nowak:** Investigation. **Paweł Michałowski:** Investigation, Writing – review & editing. **Mateusz Kowacz:** Investigation, Writing – review & editing. **Marek Schmidt:** Methodology, Resources. **Feliks Stobiecki:** Conceptualization, Methodology, Formal analysis, Writing – review & editing. **Gabriel David Chaves-O'Flynn:** Formal analysis, Writing – review & editing. **Piotr Kuświk:** Conceptualization, Methodology, Investigation, Formal analysis, Supervision, Writing – original draft, Writing – review & editing.